\documentclass[
superscriptaddress,
 amsmath,
 amssymb,
 aps,
prl,
floatfix,
twocolumn,
10pt
]{revtex4-2}
\pdfoutput=1
\usepackage[utf8]{inputenc}
\usepackage[T1]{fontenc}
\usepackage{babel}
\usepackage{graphicx}
\usepackage{amsfonts}
\usepackage{amssymb}
\usepackage{amsmath}
\usepackage{amsthm}
\usepackage{mathtools}
\usepackage{physics}
\usepackage[normalem]{ulem}
\usepackage{blindtext}
\usepackage{dsfont}

\usepackage{mathtools}
\usepackage{braket}
\renewcommand\bra[1]{{\langle{#1}|}}
\renewcommand\ket[1]{{|{#1}\rangle}}
\usepackage{graphicx}
\usepackage{bm}
\usepackage[dvipsnames]{xcolor}

\usepackage{natbib}
\setcitestyle{square,numbers}
\usepackage[colorlinks]{hyperref}
\usepackage[dvipsnames]{xcolor}
\usepackage[authormarkuptext=name,commentmarkup=uwave]{changes}
\definechangesauthor[name={J\AA L}, color=green!70!black]{ja}
\definechangesauthor[name={KS}, color=blue]{ks}
\definechangesauthor[name={MRK}, color=violet]{MRK}

\newcommand{\mrk}[1]{{\color{red} #1}}
\newcommand{\changed}[1]{{\color{black} #1}}
\setcitestyle{square,numbers}
\usepackage[colorlinks]{hyperref}
\makeatletter
\def\section#1{
    \par\emph{#1}---%
  \@ifnextchar\par\@gobble\relax}
\def\subsection#1{
    \par\emph{#1}---%
  \@ifnextchar\par\@gobble\relax}
\def\subsubsection#1{
    \par\emph{#1}---%
  \@ifnextchar\par\@gobble\relax}
\makeatother

\begin{document}

\title{Unquestionable Bell theorem for interwoven frustrated down conversion processes} 
\author{Pawe\l{} Cie\'sli\'nski}
\affiliation{International Centre for Theory of Quantum Technologies, University of Gdansk, 80-309 Gda{\'n}sk, Poland}
\affiliation{Institute of Theoretical Physics and Astrophysics, University of Gda\'nsk, 80-308 Gda\'nsk, Poland}
\author{Jan-\AA ke Larsson}

\affiliation{Institutionen f\"or Systemteknik, Link\"opings Universitet, Link\"oping, Sweden}
\author{Marcin Markiewicz}
\affiliation{International Centre for Theory of Quantum Technologies, University of Gdansk, 80-309 Gda{\'n}sk, Poland}
\affiliation{Institute of Theoretical and Applied Informatics, Polish Academy of Sciences, ul. Ba{\l}tycka 5, 44-100 Gliwice, Poland}
\author{Konrad Schlichtholz}

\affiliation{International Centre for Theory of Quantum Technologies, University of Gdansk, 80-309 Gda{\'n}sk, Poland}

\author{Marek \.Zukowski}

\affiliation{International Centre for Theory of Quantum Technologies, University of Gdansk, 80-309 Gda{\'n}sk, Poland}

\pagenumbering{arabic}

\begin{abstract}

Interwoven frustrated parametric down conversion (PDC) processes produce  interference effects based on path identity [Phys.\ Rev.\ Lett.\ 118, 080401 (2017)].  In this letter we show Bell nonclassicality of the processes: a proper violation of the Clauser-Horne inequality when the local measurements are controlled by on-off switching of the final local PDC processes. Such a non-standard approach is needed because if only the local phase shifts are used for the measurement settings, as done in the experiment reported in  Sci.\ Adv.\ 11, 1794 (2025), there exists a local realistic model of the interference, which we present in this letter. Nevertheless, the reported destructive interference is deep enough to violate the inequality when using on-off switching, so our result establishes a firm footing for non-classicality of the new interferometry, in both theory and experiment. The on-off approach to Bell analysis of path-identity-based interference forms a new platform for seeking new highly counterintuitive quantum phenomena.

\end{abstract}

\maketitle

The series of trailblazing parametric down conversion experiments by Mandel and his group culminated with a two-crystal experiment in which the interference depended on the level of path-indistinguishability of a photon, the source of which was undefined -- one of the two crystals \cite{Zou_1991}. Soon after the Zeilinger group designed and performed an experiment that they have called \textit{frustrated downconversion} \cite{Herzog_1994, HERZ-1995}. The latter was based on the idea that if we have two down conversion crystals through which the pump field passes sequentially, one can redirect the idler and signal modes emerging from the first crystal in such a way that they pass through the second crystal and overlap with phase-matched idler and signal exit modes of the second crystal. 
With that, the processes of birth of pairs of down-converted photons in the first and the second crystal can be made indistinguishable. As the same field pumps the crystals, one can have a controllable quantum superposition of the birth of the pairs in the two down-converters. Interference between the two processes, dependent on the phase shifts in idler, signal and pump field modes between the crystals can suppress or enhance the pair production. 
While the process was interesting by itself, the interference was of the first order, it could be monitored by observing just the signal or idler output of the second crystal.

In a paper entitled ``Entanglement by Path Indentity''\cite{PhysRevLett.118.080401}, new multi down conversion crystal  interferometric configurations inspired by \cite{Zou_1991} and \cite{Herzog_1994} were proposed, which utilize  fundamental indistinguishability of the sources of photons, for further generalizations see \cite{RevModPhys.94.025007}. 
The first experiment exploring these ideas \cite{Qian_2023} involves higher order interference in coincidences of two pairs of photons counted at two different measurement stations (potentially space-like separated) involving interwoven emissions from the two down conversion crystals, and measurement stations involving phase shifters and second stage parametric down conversions.
After a topological transformation, such a configuration resembles a Bell experiment. \changed{In Wang  et al. ``Violation of Bell inequality with unentangled photons'' \cite{MA}} the authors present a desk-top version of such an experiment, and claim observations of a violation of a Bell inequality. We shall argue that this is only a conjecture. Nevertheless, {\em observed destructive interference is deep enough to make it impossible for any local realistic model to explain a modified version of the experiment to be proposed here}. 

In contrast to the presentation in \cite{MA} we follow the standard Bell-type analysis, without any additional assumptions beyond those of realism, locality, and free choice of local measurement settings. 
We show that the Bell analysis in \cite{MA} rests on tacit additional assumptions. 
Note that we are not alone in this opinion, see Refs.~\cite{Price_2025,Wojcik2025}.
In our re-analysis we explicitly construct a local realistic model for the correlations which are the basis of the claims in \cite{MA} using modeling ideas \cite{LARSSON99} previously applied in \cite{Aerts1999} to give a local realistic model of Franson interferometry \cite{Franson1989}, and also in \cite{ModelNJP} to give a local realistic model of the Tan-Walls-Collett experiment \cite{TWC91}. 
Moreover, we provide an analysis which shows that the statement in \cite{MA} about ``unentangled'' photons, cannot be upheld.

The essence of our approach for showing a proper Bell inequality violation is to abandon the usual approach in optical Bell tests, in which the macroscopically controlled optical phases define the measurement settings.
In our approach, the local settings are \changed{additionally} determined by switching {\em on} or {\em off} the local pumping fields of the parametric down conversion crystals at the local measurement stations of ``Alice and Bob''. This method was inspired by the trailblazing paper by Hardy \cite{Hardy94} and recently developed at ICTQT, see Refs.~\cite{SinglephotonNJP,ModelNJP,3rdPaper,schlichtholz2023singlephoton} in which weak homodyne measurements operating in the aforementioned \textit{on-off} mode allowed to propose a version of the Tan-Walls-Collett \cite{TWC91} experiment revealing proper Bell non-classicality of a single photon, prepared in a superposition state of propagating to Alice and to Bob.

\section{Interwoven frustrated down conversion processes set-up}

\begin{figure}
    \includegraphics[width=0.8\linewidth]{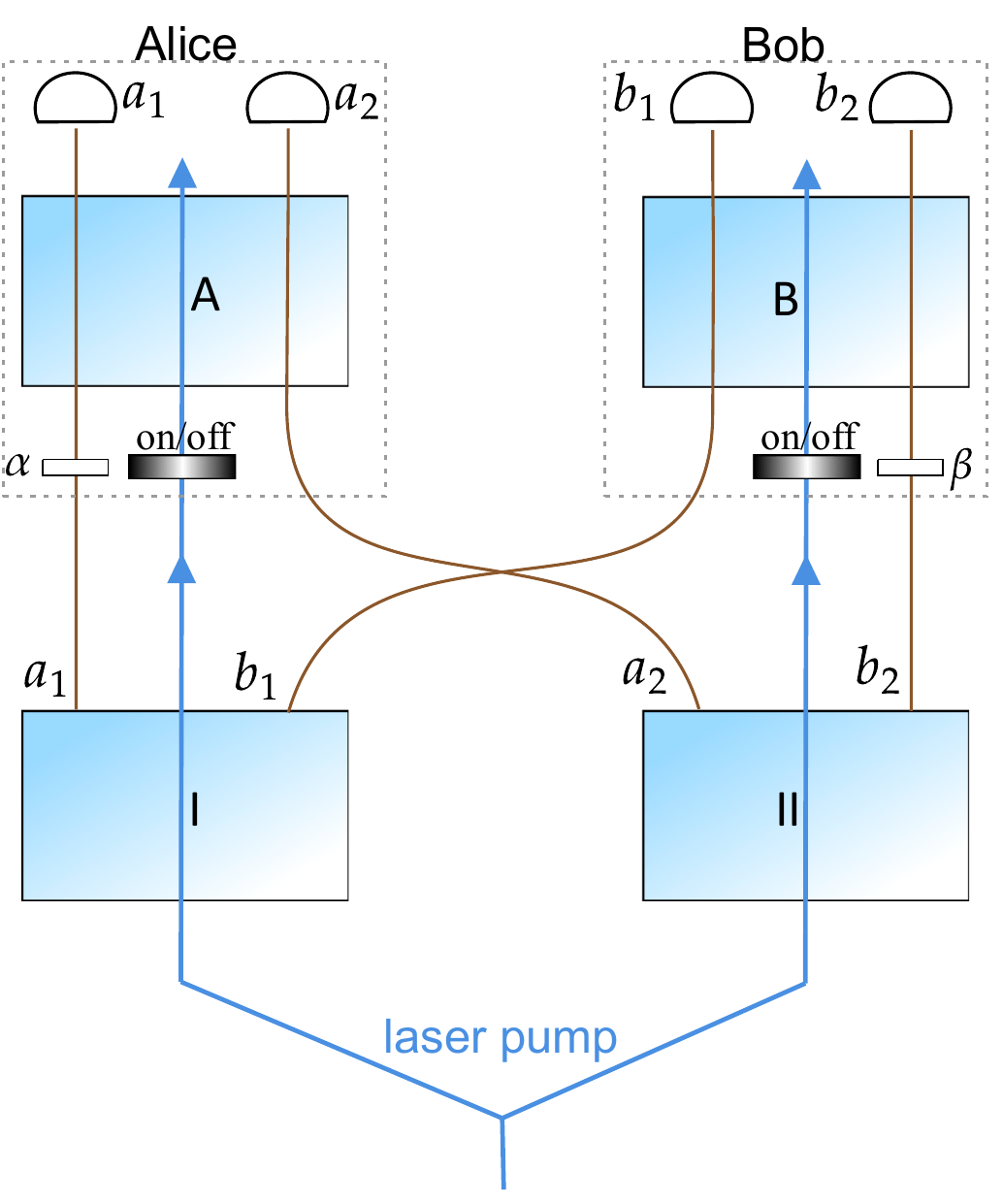}
    \caption{Schematic of the interwoven frustrated down conversion processes set-up studied in this work, both the version of \cite{MA} and our modification. Our modification contains shutters (marked by {\em on/off}), which can locally switch off the pumping fields at the Alice's and Bob's down converters $A$ and $B$. The down converters are treated as an integral part of the local measurement stations in the dotted boxes.}
    \label{fig:set-up}
\end{figure}

The schematic of the experimental set-up is given in  Fig.~\ref{fig:set-up}. The figure covers both the original version of the set-up of \cite{MA} and our modification.
The pump beam is first split into two beams, each entering the first layer of nonlinear crystals, \changed{denoted I and II and} represented by the light blue rectangles, where the parametric down conversion process occurs. The resulting output modes are then interwoven and directed toward two separate stations: Alice’s on the left and Bob’s on the right. Next, a local phase shift $\alpha$ is applied in the mode $a_1$, and $\beta$ in the mode $b_2$. The beams $a_1$ and $a_2$ enter via phase matched path the local down-converter A of Alice, while $b_1$ and $b_2$ enter the local down converter  B of Bob. In \cite{MA} the macroscopically controlled local settings are the values $\alpha$ and $\beta$ of the local phase shifts. 
The converters A and B are \changed{aligned and} phase matched in such a way  that their potential down conversion processes are into modes $a_1, a_2$ and $b_1,b_2$,  respectively. The local \changed{measurement stations} are monitoring these modes behind the crystals.
Note that the local measurement stations contain the down converters as their {\em integral} elements. 

In the experiment of \cite{MA} the phases were varied, and an interference effect was observed in double-photon counts,
    \begin{equation}\label{INTERF}
        P_{AB}(1111|\alpha,\beta)\propto [1+v\cos{(\alpha+\beta})].
    \end{equation}
Here, $1111$ stands for photon number registered in the beams $a_1a_2b_1b_2$ and $v$ for the visibility.

 \changed{The phases $\alpha,\beta$, and, crucially for our argument, the powers of local pump beams are macroscopic} parameters which affect the action of the local down converters, and are controllable by Alice and Bob. We propose to  allow Alice and Bob to switch \textit{on} or \textit{off} the local pumping fields in the measurement stations. This defines local settings of the Bell experiment.
If the phases satisfy $\alpha+\beta=\pi$ and both pumps are {\em on} we have maximal destructive interference for $1111$ counts. 

\subsection{The on-on pumping operation mode of the experiment, the basis of \cite{MA}}
For the down conversion process in the crystal with output modes $x_i$ and $y_j$ where $i,j=1,2$ and $x,y=a,b$, the Hamiltonian for the PDC process is 
   $ \hat H_{x_iy_j}=g_c(\hat x_i^{\dagger}\hat y_j^{\dagger}+h.c),$ 
where $g_c\in \mathbb{R}$ is the parameter that gives the coupling of the pump field with down-conversion modes, proportional to the field strength. Note that we use a different convention for the squeezing Hamiltonian than that of \cite{MA}; this difference has no effect on the predictions.
The above operator generates, after interaction time $t$, a unitary transformation which up to the second order in $g=g_ct$ takes the form of
\begin{equation}
    \hat U_{x_iy_j}=1+ig (\hat x_i^{\dagger}\hat y_j^{\dagger}+h.c) -\frac{g^2}{2}(\hat x_i^{\dagger}\hat y_j^{\dagger}+h.c)^2,
\end{equation}
where $\hat x_i^\dagger, \hat y_j^\dagger$ stands for creation operators in the corresponding modes. We denote the local phase shift operations $\hat U_{a_1}(\alpha)$ and $\hat U_{b_2}(\beta)$\changed{, that transform $a_1^\dagger \rightarrow e^{i\alpha}a_1^\dagger$ and $b_2^\dagger \rightarrow e^{i\beta}b_2^\dagger$, respectively.}

The final state for the shutters in \textit{on-on}  position is
\begin{equation}
\begin{split}
    &|\text{on,on};\alpha,\beta\rangle:= 
    \hat U_{a_1a_2}\hat U_{b_1b_2}\hat U_{b_2}(\beta)\hat U_{a_1}(\alpha) \hat U_{a_1b_1} \hat U_{a_2b_2}\ket{0000}\\
    &=(1 - 2 g^2) |0 0 0 0\rangle\\
    &\quad +  i g(|0 0 1 1\rangle + |1 1 0 0\rangle + e^{i \alpha} |1 0 1 0\rangle +  e^{i \beta} |0 1 0 1\rangle)\\
    &\quad-g^2(|0 0 2 2\rangle + |2 2 0 0\rangle + e^{2 i \alpha} |2 0 2 0\rangle +e^{2 i \beta} |0 2 0 2\rangle \\
    &\qquad+ [1 + e^{i (\alpha + \beta)}] |1 1 1 1\rangle +\sqrt{2} [ e^{i \alpha} |1 0 2 1\rangle  +  e^{i \alpha} |2 1 1 0\rangle  \\
    &\qquad\quad+  e^{i \beta} |0 1 1 2\rangle +  e^{i \beta} |1 2 0 1\rangle ]) +O(g^3)
  \label{eq:on-on-2}
\end{split}
\raisetag{1.4em}
\end{equation}
A cutoff at $g^2$ might be dangerous, as we are going to calculate probabilities, like the authors of \cite{MA}, of the order of $ g^4$. 
For example, the single-pair detection probability at Alice station contains apparent signaling due to the term $g^2[1 + e^{i (\alpha + \beta)}] |1 1 1 1\rangle$, because there is a correction needed from a coefficient of order $g^3$, see the Supplementary Material \cite{supplemental}.
However, the cut-off and correction do not affect the probabilities used in next section to show a proper violation of a Bell inequality.
The value of the parameter  $g$ reported in Ref.~\cite{MA} is $0.096 \pm 0.008$.

\section{Proof of proper Bell non-classicality of the interwoven frustrated down conversion process} 
 We propose a ``game rules'' for proper Bell experiments: (a) First, show that the ideal theoretical version of experiment violates a proper Bell inequality, which is based solely on  local realism. Only after a positive outcome of this investigation,  (b) discuss {\em imperfections} (loopholes) in your actual laboratory experiment, which may reduce or prohibit violation of the considered proper Bell inequality, and argue that perhaps under additional assumptions, like fair-sampling, results of your experiment suggest that local realism is not a good model for the experiment.

 We will show  Bell non-classicality of the perfect quantum predictions of the discussed modified experiment, using the  Clauser-Horne inequality~\cite{CH}, game rule (a): \begin{equation}
\label{eq:CH}
\begin{split}
CH&:=  P(A,B)+P(A,B') +P(A',B) \\ &\qquad-P(A', B')-P(A)-P(B)\leq0.
\end{split}
\end{equation}
The idea behind the modified experiment is to use the {\em on-off} local pumping fields approach for the settings, see Fig.~\ref{fig:set-up}. 
The event definitions that we use are:

\begin{description}
    \item [$A$] the pumping field at Alice's measurement station is \textit{off}; a pair of photons is registered  --- one in each detector of hers, i.e. $a_1a_2=11$.
    \item [$A'$] the pumping field at Alice's station is \textit{on}; a pair of photons is registered  $a_1a_2=11$.
    \item [$B$ {\normalfont and} $B'$]  analogous but at Bob's station, with detection $b_1b_2=11$.
\end{description}

The final evolution of the state before detections but behind all down converters and phase shifts is given below. All formulas for the states in each \textit{on-off} case are shown up to $g^2$. 

\begin{description}

\item[$(A,B)$] In the \textit{off-off}  situation the state is
\begin{equation} \label{OFF-OFF}
\begin{split}
&| \text{off,off};\alpha,\beta\rangle:=\hat U_{b_2}(\beta)\hat U_{a_1}(\alpha) \hat U_{a_1b_1} \hat U_{a_2b_2}\ket{0000}\\
    &=  (1 - g^2) |0000\rangle + i g (e^{i \alpha} |1 0 1 0\rangle   + e^{i \beta} |0 1 0 1 \rangle  )\\
    &\quad- g^2(
   e^{2 i \alpha} |2020\rangle  + 
  e^{2 i \beta}|0202\rangle  +
 e^{i (\alpha + \beta)}|1111\rangle)\\&\quad +O(g^3).
\end{split}
\end{equation}
We include the phase shifts, but they do not matter here. The state produced by the sources is $ | \text{off,off};\alpha=0,\beta=0\rangle,$ that is, a four-mode squeezed vacuum.  {\em We have neither $\ket{11xy}$ nor $\ket{xy11}$ terms  with $xy\neq 11$}. Thus,  $P(A)=P(B)=P(A,B)=g^4$. 

\item [$(A',B)$]
The \textit{on-off} situation includes the additional evolution $\hat U_{a_1a_2}$ because the down converter at Alice's station is on. This results in the state
\begin{equation}
\begin{split}
&| \text{on,off};\alpha,\beta\rangle:=  \hat U_{a_1a_2}| \text{off,off};\alpha,\beta\rangle\\
&= (1-\frac{3}{2}g^2) |0 0 0 0\rangle +i g ( |1 1 0 0\rangle  + 
   e^{i \alpha} |1 0 1 0\rangle   + 
 e^{i \beta}  |0 1 0 1\rangle)\\
 &\quad-g^2 \big( \sqrt{2} e^{i \alpha} |2 1 1 0\rangle   + e^{2 i \alpha} |2 0 2 0\rangle  + \sqrt{2} e^{i \beta} |1 2 0 1\rangle\\
 &\quad+ e^{2 i \beta} |0 2 0 2\rangle  + e^{i (\alpha + \beta)} |1 1 1 1\rangle+|2 2 0 0\rangle \big)  +O(g^3).
\end{split}
\raisetag{1.4em}
\end{equation}
Therefore, $P(A',B)=g^4.$
\item [$(A,B')$] The \textit{off-on} situation is described by the above state with $a_1a_2$ permuted with $b_1b_2$. Therefore  $P(A,B')=g^4$.
\changed{\item[$(A',B')$] The final state for the \textit{on-on} case is in the formula (\ref{eq:on-on-2}). Therefore $P(A', B')=4g^4\cos^2{\frac{1}{2}(\alpha+\beta)}=2g^4(1+\cos(\alpha+\beta))$.}
\end{description}
\changed{From this we obtain
\begin{equation} \label{VIOLATION}
    CH=g^4[-1-2\cos(\alpha+\beta)].
\end{equation}
Since  the maximum of (\ref{VIOLATION}) is $g^4$, attained for the maximally destructive interference for which $P(A', B')=0$, rule (a) is obeyed. However, the visibility imperfections in (\ref{INTERF}), which are the only ones discussed in \cite{MA}, reduce the violation of (\ref{eq:CH}). The only interference term  in the inequality is $P(A',B')$, which following (\ref{INTERF}) would be modified to $2g^4(1+v\cos(\alpha+\beta))$.  Then, the threshold visibility for violation of the inequality is $v_{\text{thr}}=1/2$. The visibility reported in \cite{MA} is $v_{\text{exp}}=0.784$, which is much larger than~0.5.}

\section{Discussion of the previously conjectured Bell non-classicality}

First, we address the claim that  the four-photon coincidence events, which are claimed to reveal Bell non-classicality, involve ``unentangled'' photons. In Ref.~\cite{MA}, page 2, the authors write  \changed{``The final state of the photons occupying the modes $a_1$, $a_2$, $b_1$, and $b_2$ by postselecting the four-photon term is''}, in our notation \changed{of their formula (1)}:
\begin{equation}
    \ket{\tilde\psi_f}=g^2(1+e^{i(\alpha+\beta)})\ket{1111}
\end{equation}
However, the above is just a component of the full quantum state. 
When post-selecting \textit{only} the $1111$ outcomes, \changed{the state reduces to $\ket{1111}$, and} one also removes the detection events, other than $1111$ that are necessary to estimate the probability of $1111$ to happen. 

 \changed{Contrary to the claim in \cite{MA} } the state on which the \changed{local measurement stations} perform {\em Bell protocol} measurements is not the state  $|\text{on,on}; \alpha, \beta\rangle$. \changed{The phase shifts $(\alpha,\beta)$ mac\-ro\-sco\-pic\-ally controlled by Alice and Bob {\em are in front of} the crystals A and B and determine the settings for the local measurements, along with the on/off power settings in our modification.

The state   before the local measurement stations is the four-mode
(weakly) squeezed vacuum $|\text{off,off}; \alpha = 0, \beta = 0 \rangle$. It is entangled. Thus, the claim in the title of \cite{MA} is unfounded, {\em cf.}  Supplemental Material \cite{supplemental} Section E.}


The other problem in \cite{MA} is that the quantum predictions for
the single-pair probabilities $P_A(11|\alpha)=P_B(11|\beta)=g^2-\frac83g^4$ are much larger than the coincident-pair probability $P_{AB}(11,11|\alpha,\beta)=2g^4[1+\cos(\alpha+\beta)]$. 
Thus, not enough coincident pairs are detected compared to single pairs to enable violation of a proper Bell inequality, compare Refs.~\cite{Pearle1970,Garg1987,Larsson1998,Aerts1999,Jogenfors2014,Larsson2014}. 
The construction, presented in \cite{MA}, of a CHSH-like inequality does not follow the standard Bell scenario in order to satisfy rule (a).
The expression called ``correlation function'' in \cite{MA} is neither the one defined by Bell and CHSH, nor the one used in probability theory and mathematical statistics, and that the inequality derived in \cite{MA} rests on three tacit additional assumptions{.} 
 An extended discussion of the issue is given in the Supplementary Material.

\subsection{Local realistic model for the Wang et al.~experiment: no Bell theorem with unentangled photons}

To make this completely explicit, we construct a local hidden variable model that reconstructs all of the observed statistics. That is
\begin{equation}
\begin{split}
    &P_A(11|\alpha)=P_B(11|\beta)=g^2 -\frac{8}{3}g^4,\\
    &P_{AB}(11,11|\alpha,\beta)=2g^4[1+\cos(\alpha+\beta)]
    \label{eq:model1}
\end{split}
\end{equation}
where the local probabilities are calculated using the expansion of the state (\ref{eq:on-on-2}) up to $g^4$ to avoid the apparent signaling, see the Supplementary Material \cite{supplemental}. Note that the above probabilities exactly reproduce the quantum predictions for all relevant quantum
events in the experiment considered in \cite{MA}. Therefore, 
any possible relation with the $\pm1,\pm 1$ “outcomes” defined
in \mrk{\cite{MA}} is irrelevant here, see the discussion in Supplementary Material \cite{supplemental}.

We assume that the system has hidden variables $\lambda_A = \lbrace \phi_A, r_A \rbrace$ and similar for $\lambda_B$. We let $\phi_A + \phi_B = 2\pi$, and $r_A + r_B = 1$, where $\phi_A$ is uniformly distributed over $[0, 2\pi]$ and $r_A$ is uniformly distributed over the unit interval. The relation between $\lambda_A$ and $\lambda_B$ can exist because \emph{a)} the down converters share a pump, and \emph{b)} they also share one beam each from the ``remote'' initial down conversion crystal, either of these suffice. For
reasons of locality they do not, however, have access to the ``remote'' interferometer phase delay $\alpha$ or $\beta$. 

The  model works as follows. Alice's measurement station has a local parameter $\alpha$, and a photon pair is detected if $\sin(\phi_A+\pi-\alpha)\geq 0$ and either $r_A \leq c \sin(\phi_A+\pi-\alpha)$ or $r_A\geq 1-d$. The introduced parameters $c$ and $d$ are constants to be chosen \changed{so that the model gives the probabilities of Eqn.~(\ref{eq:model1}), and this requires} that the above intervals do not overlap, meaning $c+d\le1$. The local probability for Alice is now given as
\begin{equation}
\begin{split}
    P_A(11|\alpha)
    &=\int_{\alpha-\pi}^{\alpha} \frac{d\phi_A}{2 \pi} [c \sin(\phi_A+\pi-\alpha)+d]
    \\&
    =\frac{c}{\pi}+\frac{d}{2}.
    \end{split}
\end{equation}
At Bob's station, a photon pair is detected if $\sin(\phi_B - \beta) \geq 0$ and either $r_B \leq c \sin(\phi_B - \beta)$ or $r_B \geq 1-d$, which makes $P_B(11|\beta)=P_A(11|\alpha)$. 
\begin{figure}
    \centering
    \includegraphics[width=\linewidth]{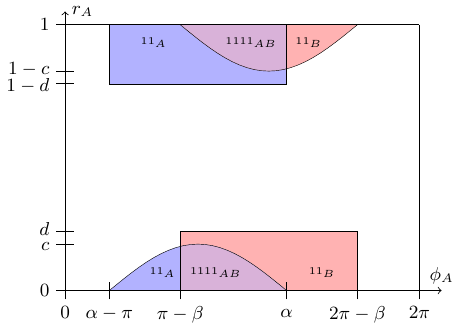}
    \caption{Local hidden variable model that reproduces the statistics used in Ref.~\cite{MA}.  Overlap area for double pair detection. Blue area represents a pair detection at A, and red area a pair detection at B. The purple area shows the four-detection at A and B.}
    \label{fig:joint_probability}
\end{figure}%

The probability of $P_{AB}(11,11|\alpha, \beta)$ of double pair detection is calculated as follows. From $\phi_A+\phi_B=2\pi$ one obtains that the region of interest for $\phi_A$ to  simultaneously fulfill conditions  $\sin(\phi_A+\pi-\alpha)\geq 0$ and $\sin(\phi_B-\beta)\geq 0$ is given as the intersection between  $\beta+\pi \leq \phi_A \leq 2 \pi - \beta$ and $\alpha-\pi \leq \phi_A \leq \alpha$, see Fig.~\ref{fig:joint_probability}.
When $c\le d$, in the case of $0\leq\alpha-\pi\leq \pi-\beta \leq \alpha$ the intersection of the local 11 events, see Fig.~\ref{fig:joint_probability}, is  
\begin{equation}
\begin{split}
    P_{AB}(11,11|\alpha,\beta)&=2\int_{\pi-\beta}^{\alpha} \frac{d \phi_A}{2 \pi} c \sin(\phi_A+\pi-\alpha) 
    \\&
    =\frac{c}{\pi}[1+\cos(\alpha+\beta)]
\end{split}
\end{equation}
It is straightforward to check that other values of $\alpha$ and $\beta$ give the same formula. 
By choosing
$c=2\pi g^4$ and $d=2 g^2 -\frac{28}{3}g^4,$
we obtain
exactly the probabilities in {Eqn.~}(\ref{eq:model1}). 
This choice of $c$ and $d$ gives us $c+d\le1$, and $c\le d$ if
$g\leq \sqrt{6/(6\pi+28)}\approx 0.358$ 
which covers  $g=0.096$ as  reported in \cite{MA}, and in practice the whole range of physical parameters $g$ for which our approximated PDC process calculations are generally valid.

The remaining events predicted with the corrected approximate state do not show any interference, and are therefore trivial to add to the model.

\section{Conclusions}

After our modification as shown in Fig.~\ref{fig:set-up} the set-up for interwoven frustrated down conversion processes used in \cite{MA} can reveal proper Bell non-classicality.  The violation is due to the entanglement of the four-mode squeezed vacuum state distributed to the local measurement stations. The macroscopically controlled settings of the local measurement apparatuses, of which an integral part are the down converters of Alice and Bob, are in the case of experiment of \cite{MA}  phases $\alpha$ and $\beta$, and \changed{in our modification additionally} the amplitudes of the pumping fields entering the crystals A and B of the measurement stations. Our modification {of the arrangement }constitutes a new generalized all-optical measurement scheme employing non-linear optics in Bell experiments -- a step forward with respect to a, e.g., weak homodyne detection \cite{TWC91,Walmsley20}, which is based purely on linear optical devices.

Our local hidden variable model for the ideal quantum predictions for the correlations used in \cite{MA} shows that the CHSH-like inequality of \cite{MA} is not a proper Bell inequality. Therefore, the claims in \cite{MA} about revealing Bell non-classicality by only manipulating the local phases are just conjectured. The analysis contains tacit assumptions that create loopholes that are impossible to close by using better laboratory equipment, better alignment, etc. However, as we have shown they can be removed by re-arranging the measurement scheme to the on-off one.
As the destructive interference reported in \cite{MA}  is deep enough to violate the inequality, the non-classicality of the new interferometry gains a firm footing in theory and laboratory.

\section{Acknowledgments}
This work is supported by project no. FENG.02.01-IP.05-0006/23, financed by the (IRAP) MAB FENG program 2021-2027, Priority FENG.02, Measure FENG.02.01., with the support of the FNP (Foundation for Polish Science). PC acknowledges the support of FNP within the START programme.
J{\AA}L acknowledges support from the Swedish Science Council project no 2023-05031.

\nocite{Bell.64}\nocite{CHSH}\nocite{Streater2007}
\bibliography{biblioFRUfinal}

\pagebreak
\textbf{Supplemental Material}\\
\setcounter{equation}{0}
\renewcommand{\theequation}{S\arabic{equation}}
\setcounter{figure}{0}
\renewcommand{\thefigure}{S\arabic{figure}}
This supplemental material is divided into five sections. Section A discusses the problem of building ``CHSH inequalities'' in Wang et al. \cite{MA} using non-standard correlation functions. Section B presents additional hidden assumptions over the standard Bell scenario used in the CHSH inequality derivation performed by Wang et al. Section C discusses how local models can exploit absence of these assumptions to violate this inequality. Section D presents how one should perform cut-off of the  state expansions with respect to the coupling constant $g$ from the main text, in  order to correctly reproduce all probabilities in the experiment and avoid apparent signaling.  Section E Shows that the observed interference cannot to ascribed to "unentangled photons".

\section{A: The ``correlation function'' and ``CHSH inequalities'' in Wang et al. differ form the Bell-CHSH ones}

Here we analyze the description of the authors of \cite{MA} of how they construct an inequality which they interpret as a version of the CHSH one. We refer directly to the text of \cite{MA}, and thus the Reader should directly compare our considerations here with the formulas and statements of Ref.~\cite{MA}.


The correlation function used by Bell \cite{Bell.64}, CHSH \cite{CHSH} and many others, specific for dichotomic numerical values of  outcomes $\pm1$, obtained from standard probability calculus, reads:
\begin{equation} \label{Bell-correlation}
    E(\alpha, \beta)= \sum_{j=\pm1}\sum_{i=\pm1} ijP(i,j|\alpha, \beta),
\end{equation}
where $i$ and $j$ are numbers which represent the measured (obtained) values, and $\alpha$ and $\beta$ are the settings of the respective local measuring apparatuses.  As we discuss here the contents of the paper \cite{MA}, the symbols for the settings in Bell experiment that  we use are $\alpha$ and $\beta$, but of course the formula is general. The outcomes $i,j$ and parameters $\alpha,\beta$ are macroscopic and directly accessible in the experiments. 
We stress that, in a proper Bell  analysis 
macroscopic setting parameters  $\alpha,\beta$ are {\em fixed} when calculating the correlation function, and the outcomes $i,j$ are not chosen by the experimenter. 

In the paper \cite{MA}, the authors give an unconventional definition of ``probability'' which is used in their attempted version of CHSH inequality:
\begin{equation} \label{paper-2}
\begin{split}
    &P_{paper}(r,s|\alpha , \beta)\\
    &=
    \frac{\mathcal N(+1,+1|\alpha +\frac{1-r}{2}\pi, \beta+ \frac{1-s}{2}\pi )}{\sum_{r'=\pm1}\sum_{s'=\pm1} \mathcal N(+1,+1|\alpha +\frac{1-r'}{2}\pi, \beta+ \frac{1-s'}{2}\pi)},
\end{split}
\end{equation}
where $\mathcal{N}$ stands for the number of double-pair events and $r=\pm1$, $s=\pm1$ are parameters that control presence or absence of a local phase shift of $\pi$ radians. 
We will first attempt to analyze these without extra assumptions beyond local realism, and return to the extra assumptions below.
We especially note that the \textit{parameters} $r$ and $s$ are not measurement outcomes, so Eqn~(\ref{paper-2}) does not have the form of a standard (conditional) probability. 

The events counted in the denominator are all events in which a 1111 coincidence occurs at the different possible combinations of macroscopically controllable settings $\alpha$, $\alpha+\pi$ and $\beta$, $\beta+\pi$, and this is perhaps to be expected since the authors of \cite{MA} claim that they postselect these events. 
The form of the ``probability'' (\ref{paper-2}) therefore suggests it is a conditional probability, conditioned on the 1111 coincidences. 
The different events contained in the denominator contain different values of the parameters $r$ and $s$, so the expression calculates the conditional probability that the \textit{parameters} $r$ and $s$ have a specific value; in other words the probability that the local phase is shifted by $0$ or $\pi$.
This is under a tacit assumption that local experimenters choose between these shifts randomly, without a bias.
With the use of this assumption Eqn.~(\ref{paper-2}) indeed forms a conditional probability which, to be precise, should be denoted 
\begin{equation} \label{paper-2b}
\begin{split}
&P_{paper}(r,s|1111,\alpha , \beta, r=\pm 1, s=\pm 1)\\
    &=
    \frac{\mathcal N(+1,+1|\alpha +\frac{1-r}{2}\pi, \beta+ \frac{1-s}{2}\pi )}{\sum_{r'=\pm1}\sum_{s'=\pm1} \mathcal N(+1,+1|\alpha +\frac{1-r'}{2}\pi, \beta+ \frac{1-s'}{2}\pi)}\\ 
    &=
    \frac{\mathcal N(1111,\alpha,\beta,r,s)/P(\alpha,\beta,r,s)}{\sum_{r'=\pm1}\sum_{s'=\pm1}  \mathcal N(1111,\alpha,\beta,r',s')/P(\alpha,\beta,r',s')}\\
    &=\frac{\mathcal N(1111,\alpha,\beta,r,s)}{\mathcal N(1111,\alpha,\beta,r=\pm1,s=\pm1)}\\
\end{split}
\end{equation}

But the local phase is an external input; the values of $r$ and $s$ are chosen by the experimenters. 
Calculating the conditional probability of parameter values conditioned on the measurement outcomes is akin to the well-known ``prosecutor's fallacy'' from statistics, see e.g. \cite{Streater2007}.
A proper analysis should instead estimate the probability of measurement outcomes conditioned on the input parameter values; thus we need to use the counts of opposite \textit{outcomes} under the same experimental \textit{settings}. We now turn to how the authors of \cite{MA} claim to achieve this.


\section{B: Necessary additional assumptions for CHSH inequality derivation performed in Wang et al.}

To avoid the mentioned problems Ref.~\cite{MA} uses three assumptions (in addition to local realism) to derive the CHSH inequality in their setup. 
In the text preceding Ref.~\cite{MA} Eqn~(3) they state that they use the outcome value $+1$ to denote local pair detection, and then \textit{also use the outcome value $-1$} for the setting $\alpha$, $\beta$, assuming that
\begin{subequations}
\begin{eqnarray}
    N(+1,-1|\alpha,\beta)&=&
    N(+1,+1|\alpha,\beta+\pi)\\
    N(-1,+1|\alpha,\beta)&=&
    N(+1,+1|\alpha+\pi,\beta)\\
    N(-1,-1|\alpha,\beta)
    &=&N(+1,+1|\alpha+\pi,\beta+\pi)
\end{eqnarray}
\label{eq:unobservedp}
\end{subequations}
One should stress that this is in actual fact two assumptions:
\begin{enumerate}
    \item there exists an unobserved local outcome $-1$ at each detection station, and
    \item the probabilities of the assumed outcomes have the symmetry of Eqns.~(\ref{eq:unobservedp}).
\end{enumerate}
To motivate assumption 2, the authors of \cite{MA} write 
\begin{quote}
    In particular, the coincidence count under a $\pi$ phase shift, $N(+1,+1 | \alpha + \pi,\beta)$, is complementary to the count in the original setting, $N(+1,+1 | \alpha,\beta)$, and therefore can be identified as $N(-1,+1 | \alpha,\beta)$, which is also complementary to $N(+1,+1 | \alpha,\beta)$.
\end{quote}
If one were to speculate, it would seem that the authors' intuition is guided by the behavior of a spin-$\frac12$ measurement: measuring first at planar angle $\alpha$ and then at angle $\alpha+\pi$ would give opposite outcomes.
If the first gives outcome $+1$ the second gives outcome $-1$ and vice versa.
Note that the two spin measurements are not complementary, they are compatible. 
This seems to be the reasoning behind the authors of Ref.~\cite{MA} assuming the symmetry (\ref{eq:unobservedp}).

However, in the setup of \cite{MA} the coincidence measurement does not even have a local $-1$ outcome, in fact it does not correspond to an eigenvalue of a quantum observable. 
Its existence must be postulated. 
Indeed, one must also postulate that the probabilities associated with the unobservable $-1$ outcome has the desired symmetry~(\ref{eq:unobservedp}).
Assuming such a symmetry for an unobserved, in fact unobservable, outcome is unwarranted.
Only if one accepts this unwarranted assumption does Eqn.~(\ref{paper-2}) transform into something of the form of
the standard conditional probability distribution:
\begin{equation}
    P(r,s|\alpha,\beta)=
    \frac{\mathcal N(r,s|\alpha, \beta)}{\sum_{r'=\pm1}\sum_{s'=\pm1} \mathcal N(r',s'|\alpha, \beta)}.
    \label{eq:condp}
\end{equation}

Even at this point the so constructed conditional probabilities cannot be directly used to violate the CHSH inequality, in contrast to what is claimed in Ref.~\cite{MA}.
Violation is only possible if either the denominator is close enough to $\mathcal N_\text{tot}$, the total repetition count of the experimental runs in the setup including those that end with vacuum, or if assuming that the postselected subset is a fair sample of all possible outcomes \cite{Pearle1970,Garg1987,Larsson1998,Aerts1999,Larsson2014}.
Under assumptions 1-2 we find that the denominator is proportional to $g^4\mathcal N_\text{tot}\ll \mathcal N_\text{tot}$ both in the range where our approximate state calculation is valid, and in experiment.
This is far below what is needed for a proper violation, so the postselection prohibits violation, see the discussion in the main text.

The remaining option is to assume that the postselected subset is a fair sample of all possible outcomes. The term ``fair sampling'' is used for a similar assumption in the context of the Bell Theorem, to address imperfections in real laboratory experiments, related with the detection inefficiency. The detected particle counts are treated as a fair statistical sample of all potential results, even though many results are lost due to inefficient detection. The authors of \cite{MA} instead assume that the postselection they perform is fair even without experimental imperfections (in fact, even for the unobservable outcome $-1$). We have therefore elected to use the term ``fair postselection'' for this particular version of the assumption.
Thus, to obtain a violation of local realism, the experiment in \cite{MA} needs three additional assumptions, beyond local realism, namely
\begin{enumerate}
    \item Existence of an unobservable $-1$ local outcome.
    \item Outcome probabilities have the symmetry of Eqns.~(\ref{eq:unobservedp}).
    \item Fair postselection.
\end{enumerate}

\section{C: Consequences if not using the three extra unwarranted assumptions}
The LHV model constructed in the main text obeys none of the  three assumptions identified in Section B, but can be extended to obey  assumption 1 and 2 as follows.
Label the ``pair detection'' outcome ``$+1$'' and recall that this outcome occurs if the hidden variables are such that $\sin(\theta_A+\pi-\alpha)\ge0$ and either $r_A\le c\sin(\phi_A+\pi-\alpha)$ or $r_A\ge1-d$. 
Now add a second outcome $-1$ and add this to the model by letting it occur if the hidden variables are such that $\sin(\theta_A-\alpha)\ge0$ and either $r_A\le c\sin(\phi_A-\alpha)$ or $r_A\ge1-d$. 
And similar for the second site, along the lines of the main text. 

\begin{figure}
    \centering
    a)\includegraphics[width=\linewidth]{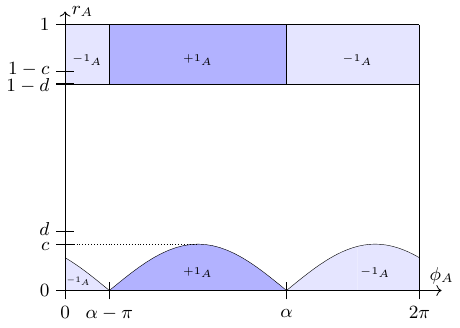}\\
    b)\includegraphics[width=\linewidth]{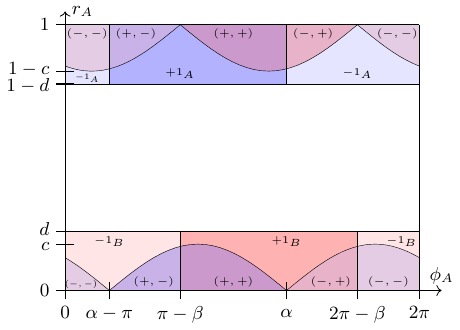}
    \caption{Model with $-1$ outcome added, for which probabilities are symmetric. a) Outcome areas for the detector at A. b) Overlap areas for different outcome combinations. Blue areas represent $\pm1$ outcomes at A, and red areas $\pm1$ outcomes B. The different shaded purple areas show the coincidences at A and B.}
    \label{fig:joint_probability_added_outcome}
\end{figure}

This gives identical $P(+1,+1|\alpha,\beta)$ as the model in the main text, and adds a $-1$ outcome (assumption 1) with equal probabilities for the different possible outcome pairs (assumption 2).
In fact, any $\lambda_A=\{\phi_A,r_A\}$ that gives outcome $+1$ for setting $\alpha+\pi$ will also gives outcome $-1$ for setting $\alpha$ and vice versa, see Fig.~\ref{fig:joint_probability_added_outcome}.
This model does not obey fair postselection (assumption 3) because when the setting changes, the postselected ensemble changes: hidden variable points $\lambda_A$ that contribute a coincidence for the setting $\alpha,\beta$ may not do so for different settings.
If $c=d=\tfrac12$ the denominator of Eqn.~(\ref{eq:condp}) equals $\frac2\pi N_\text{tot}$ \cite{LARSSON99}, far beyond the region where the approximate state is useful and far beyond what was reached in experiment.

\begin{figure}
    \centering
    a)\includegraphics[width=\linewidth]{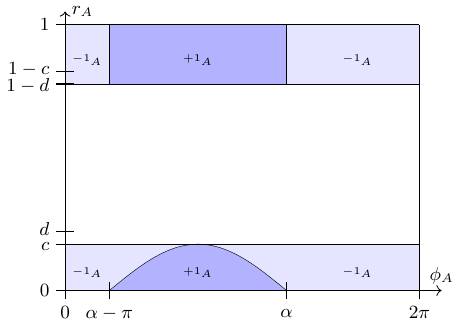}\\
    b)\includegraphics[width=\linewidth]{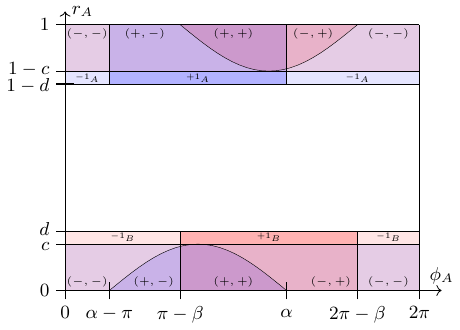}
    \caption{Model with $-1$ outcome added, that obeys fair postselection. a) Outcome areas for the detector at A. b) Overlap areas for different outcome combinations. Blue areas represent $\pm1$ outcomes at A, and red areas $\pm1$ outcomes B. The different shaded purple areas show the coincidences at A and B.}
    \label{fig:joint_probability_fair_sampling}
\end{figure}

It is also possible to construct an LHV model that obeys assumption 1 and 3 as follows (note that assumption 2 needs assumption 1 so cannot be independently obeyed). 
Change the above construction by instead adding the $-1$ outcome to the model by letting it occur if the hidden variables are such that $c\sin(\theta_A+\pi-\alpha)\le r_A\le c$ or [$\sin(\theta-\alpha)>0$ and $r_A\ge1-d$]. 
And similar for the second site, along the lines of the main text. 
This \textit{also} gives identical $P(+1,+1|\alpha,\beta)$ as the model in the standard text, and adds a $-1$ outcome (assumption 1).
Also here, any $\lambda_A=\{\phi_A,r_A\}$ that gives outcome $+1$ for setting $\alpha+\pi$ will also gives outcome $-1$ for setting $\alpha$, see Fig.~\ref{fig:joint_probability_fair_sampling}, but the converse is not true, so this model does not obey the symmetry condition (assumption 2). The model \textit{does} obey fair postselection (assumption 3) because when the setting changes, the postselected ensemble stays the same: hidden variable points $\lambda_A$ that contribute a coincidence for some setting $\alpha,\beta$ will do so for all possible settings. 
In this model if $c=d=\tfrac12$ the denominator of Eqn.~(\ref{eq:condp}) equals $N_\text{tot}$, so without assumption 2 the outcomes can be reproduced with an LHV model even with fair postselection, or indeed with no postselection and 100\% coincidences.

All of this, taken together, implies that \textit{all three assumptions} are crucial to derive the CHSH inequality used in Ref.~\cite{MA}.

\section{D: On the necessary correction of the expansion of the unitary operators needed for proper calculation of the probabilities}
Here, we will discuss the problem of the cut-off of the state $\ket{\text{on,on}; \alpha,\beta}$ up to second order in $g$ (Eq. (3) in the main text) and that it is insufficient to properly calculate some probabilities in the experiment. This is because the probabilities for which one can observe the interference  effect are of order $g^4$.  Thus, approximation to second order in amplitudes is unjustified, as corrections to amplitudes of order $g^4$ can also contribute to the probabilities of specific events at this order. Thus, to obtain probabilities valid up to $g^4$ one needs to consider Taylor expansion of operator $\hat U$ up to order $g^4$. 
However, one can simplify the calculation, as only very specific terms can contribute to probabilities in the considered order. More specifically, terms proportional to $g^4$ are relevant only if they provide correction to amplitude of order $g^0$. This means that for the considered state $\ket{\text{on,on}; \alpha,\beta}$, only correction to the vacuum state is relevant. More explicitly, the amplitude for $\ket{0000}$ after correction is given by $1-2g^2+g^4(7/3+\exp[i(\alpha+\beta)])$ which results in correction to probability of event of no click $g^4(14/3+\cos(\alpha+\beta))+O(g^6)$. Similarly, corrections to amplitudes of order  $g^3$ are relevant only when dominant part of the amplitude is of order $g^1$. This leaves only corrections to states $\ket{1100},\,\ket{0011},\,\ket{1010},\,\ket{0101}$ being relevant. The full relevant correction to the $\ket{\text{on,on}; \alpha,\beta}$ state is:
\begin{equation}
\begin{split}
    &\Bigl(\frac{7}{3}+e^{i(\alpha+\beta)}\Bigr)g^4\ket{0000}
-ig^3\frac{10}{3}\left(e^{i\alpha}\ket{1010}+e^{i\beta}\ket{0101}\right)\\
    &-ig^3\Bigl(\frac{7}{3}+e^{i(\alpha+\beta)}\Bigr)(\ket{1100}+\ket{0011}).
    \end{split}\raisetag{1.5em}
\end{equation}
Note that this correction gives us three extra interference terms  not present in the calculation using the expansions of the unitary operators up to the second order in $g$. 

This turns out to be extremely important for the consistency of the description of the (family of) states $\ket{\text{on,on}; \alpha,\beta}$. Due to the sharp cutoff approximation, it seems that one can have signaling. This appears in two ways. First, one can calculate using (3) the local probability of observing the event of detecting one photon in each detector of Alice. It reads $P_A(11)=g^2+2g^4(1+\cos(\alpha+\beta))$. This depends on the remote setting of the second party, $\beta$. However, due to the interference term which appears due to including correction to $\ket{1100}$ (with the adjusted amplitude of $ig -ig^3[7/3+e^{i(\alpha+\beta)}]$), the corrected $P_A(11)=g^2-\frac{8}{3}g^4+O(g^6)$ becomes no-signaling in the expected order  of the approximation. The second problem appears as the normalization factor for the state (3) determined by the norm square of the state $N^2=1 + 18 g^4 + 2 g^4 \cos(\alpha+\beta)$ is also modulated by the sum of the phases. This again is circumvented by the included correction, as then $N^2=1+O(g^6)$. 

The relevant corrections from $g^4$ order for the remaining settings are the following: for off-off state:
\begin{equation}
    \frac{2}{3}g^4\ket{0000}-ig^3\frac{4}{3}\left(e^{i\alpha}\ket{1010}+e^{i\beta}\ket{0101}\right),
\end{equation}
for on-off:
\begin{equation}
    \frac{11}{8}g^4\ket{0000}-ig^3\frac{7}{3}\left(e^{i\alpha}\ket{1010}+e^{i\beta}\ket{0101}\right)-ig^3\frac{11}{6}\ket{1100},
\end{equation}
for off-on
\begin{equation}
    \frac{11}{8}g^4\ket{0000}-ig^3\frac{7}{3}\left(e^{i\alpha}\ket{1010}+e^{i\beta}\ket{0101}\right)-ig^3\frac{11}{6}\ket{0011},
\end{equation}
Note that for the off-off case, local probabilities $P(A)$ and $P(B)$, which are used in the main text for calculating the violation of the Clauser-Horne inequality, are not affected by these corrections.

In general, the proper truncations for the calculations of the probabilities can be summarized as follows. For any unitary operator {$\hat U(g)=e^{ig\hat{H}}$, with obviously  Hermitian $\hat H$,}  its truncated version $\hat U_T(g)$ obtained by Taylor expansion up to order $g^n$ gives proper expectation values of any observable $\hat O$ with accuracy up to also $g^n$. Taylor series expansion of  expectation values are also preserved, that is 
\begin{equation}
    \bra{\psi}\hat U^\dagger(g)\hat O\hat U(g)\ket{\psi}-\mathcal{T}_{n}\bigl(\bra{\psi}\hat U_T^\dagger(g)\hat O\hat U_T(g)\ket{\psi}\bigr)=O(g^{n+1}),
\end{equation}
where $\mathcal{T}_{n}()$ denotes the Taylor expansion up to the $n$-th order.
For the norm of the transformed state one has  $\bra{\psi}\hat U_T^\dagger(g)\hat U_T(g)\ket{\psi}=1+O(g^{n+1})$. Thus, when calculating probabilities up to $n$-th order approximation, one does not need to renormalize the transformed state.

\section{E: the interference is due to entanglement in the initial state}
The initial squeezed state, after the phase shifts reads (this is formula (5) of the main text):
\begin{equation} \label{SQUEEZ}
\begin{split}
&| \text{off,off};\alpha,\beta\rangle:=\hat U_b(\beta)\hat U_a(\alpha) \hat U_{a_1b_1} \hat U_{a_2b_2}\ket{0000}\\
    &=  (1 - g^2) |0000\rangle + i g (e^{i \alpha} |1 0 1 0\rangle   + e^{i \beta} |0 1 0 1 \rangle  )\\
    &\quad- g^2(
   e^{2 i \alpha} |2020\rangle  + 
  e^{2 i \beta}|0202\rangle  +
 e^{i (\alpha + \beta)}|1111\rangle)\\&\quad +O(g^3).
\end{split}
\end{equation}
It is clear that mere registration of the photons (pumps off in crystals A and B) would not lead to any interference --- simply because moduli of all amplitudes do not depend on the phases. However if the pumps in the crystals are switched on the following happens. The description of the experimental situation evolves to $\hat U_{a_1a_2}\hat U_{b_1b_2} |\text{off,off};\alpha,\beta\rangle$. The action of these unitary transformations give an additional term that contains $\ket{1111}$ only in the case of the first component  $(1 - g^2) |0000\rangle$. This additional term reads $-g^2(1 - g^2) |1111\rangle $, and approximated to the second power of $g$ it reads $-g^2 |1111\rangle $. The last relevant component in the initial expansion (\ref{SQUEEZ}),   within the approximation to the second power of  $g$, is $-g^2e^{i (\alpha + \beta)}|1111\rangle$. Within the approximation to $g^2$ it  is not affected by the transformations. Thus, combining the two terms we get 
\begin{equation}
-g^2[1+e^{i (\alpha + \beta)}]|1111\rangle.    
\end{equation}
This is the term of the expansion of $|\text{on,on};\alpha,\beta\rangle$ responsible for the observation of the interference in $1111$ counts, when pumps in crystals A and B are on. Note that all that is possible only due to the fact that within (\ref{SQUEEZ}),  before the phase shifts. i.e. with $\alpha=\beta=0$ we have an {\em entangled} quantum superposition
\begin{equation}
   (1 - g^2) |0000\rangle-g^2|1111\rangle
\end{equation}
which is a four mode {\em asymmetric} GHZ state.
A mere probabilistic mixture of $\ket{0000}$
and $\ket{1111}$ would never lead an interference in $\ket{1111}$ within the considered experimental setup of Fig 1. 

Thus, one cannot claim that the observed interference is due to ``unentangled photons''.
Since the experiment uses states of modes of optical fields, which are superpositions of components with differing total numbers of photon excitations and even vacuum, one should rather talk here about mode entanglement.

\end{document}